\documentclass[journal]{IEEEtran}

\ifCLASSINFOpdf
\else
   \usepackage[dvips]{graphicx}
\fi
\usepackage{url}

\ifCLASSOPTIONcompsoc
\usepackage[caption=false,font=normalsize,labelfont=sf,textfont=sf]{subfig}
\else
\usepackage[caption=false,font=footnotesize]{subfig}
\fi

\usepackage{graphicx}
\usepackage{amsmath}
\usepackage{amsfonts}
\usepackage[utf8]{inputenc}
\usepackage[T1]{fontenc}
\usepackage[absolute,overlay,showboxes]{textpos}

\newcommand{\M}[1]{\boldsymbol{\mathbf{#1}}}

\begin{document}

\TPMargin{1mm}
\begin{textblock*}{20cm}(1cm,26.6cm)
\scriptsize
\noindent
\copyright 2019 IEEE.  Personal use of this material is permitted.  Permission from IEEE must be obtained for all other uses, in any current or future media, including reprinting/republishing this material for advertising or promotional purposes, creating new collective works, for resale or redistribution to servers or lists, or reuse of any copyrighted component of this work in other works.\\
This is the author's version of an article that has been published in this journal. Changes were made to this version by the publisher prior to publication. The final version of record is available at \url{http://dx.doi.org/10.1109/LSP.2019.2929440}
\end{textblock*}

\title{Audio Source Separation Using Variational Autoencoders and Weak Class Supervision}

\author{Ertuğ Karamatlı, Ali Taylan Cemgil, \IEEEmembership{Member, IEEE}, and Serap Kırbız
\thanks{This work was supported by TÜBİTAK under grant 215E076.}
\thanks{E. Karamatlı is with the Department of Computer Engineering, Boğaziçi University, İstanbul, Turkey and sahibinden.com, İstanbul, Turkey (e-mail: ertug@karamatli.com).}
\thanks{A. T. Cemgil is with the Department of Computer Engineering, Boğaziçi University, İstanbul, Turkey (e-mail: taylan.cemgil@boun.edu.tr).}
\thanks{S. Kırbız is with the Department of Electrical and Electronics Engineering, MEF University, İstanbul, Turkey (e-mail: kirbizs@mef.edu.tr).}}

\markboth{Journal of \LaTeX\ Class Files, Vol. 14, No. 8, August 2015}
{Shell \MakeLowercase{\textit{et al.}}: Bare Demo of IEEEtran.cls for IEEE Journals}
\maketitle

\begin{abstract}
In this paper, we propose a source separation method that is trained by observing the mixtures and the class labels of the sources present in the mixture without any access to isolated sources. Since our method does not require source class labels for every time-frequency bin but only a single label for each source constituting the mixture signal, we call this scenario as weak class supervision. We associate a variational autoencoder (VAE) with each source class within a non-negative (compositional) model. Each VAE provides a prior model to identify the signal from its associated class in a sound mixture. After training the model on mixtures, we obtain a generative model for each source class and demonstrate our method on one-second mixtures of utterances of digits from 0 to 9. We show that the separation performance obtained by source class supervision is as good as the performance obtained by source signal supervision.
\end{abstract}

\begin{IEEEkeywords}
Weak Supervision, Source Separation, Variational Autoencoders
\end{IEEEkeywords}

\IEEEpeerreviewmaketitle

 \vspace{-0.4cm}
\section{Introduction}
Non-negative matrix factorization (NMF) \cite{Lee2001,SmaragdisBrown2003} has been a popular approach for audio source separation which
decomposes a magnitude spectrogram of the mixture signal into additive parts-based decompositions named as basis functions. More recently, deep learning approaches achieved state-of-the-art results in this domain \cite{Wang2018}. They are commonly used in a supervised and discriminative setting together with time-frequency masking \cite{Huang2015}. One approach \cite{Grais2017}, which we also make use of in our work, is to use denoising autoencoders (DAEs) that take the mixture signal as input and produce a reconstruction of each target source signal. Such models usually require a large training set of isolated ground truth source signals in order to be effective which may be hard to come by in some circumstances. Weak supervision is a form of supervised learning where the targets contain information at a higher level of abstraction than needed for the task at hand and/or they contain noise. In \cite{Zhang2018}, a weakly supervised source separation method is proposed that does not require a correspondence between the separated source signals and its mixture during training. Our work is related to informed source separation \cite{Liutkus2013}, more specifically, score-informed source separation \cite{Ewert2012} for music signals where instrument activation through time is used to guide the separation of instrument sounds. In \cite {Ewert2017}, class activity penalties and structured dropout are used for score-informed source separation by applying constraints to the latent units of an autoencoder (AE). In \cite{Sobieraj2018}, an NMF method is proposed that is trained on weakly labeled data. Another work that utilizes class information is \cite{Kameoka2018} where a conditional variational autoencoder (VAE) is trained as a universal generative model to represent known source classes. The model is then used for semi-blind source separation. One related work is the joint separation and classification of sound events \cite{Kong2018} where authors not only classify the sound events but also localize them in a time-frequency representation by using only weak labels. In \cite{Hershey2016}, a clustering based method for discriminative embeddings are proposed.

In this paper, we investigate a setup where we do not have the source signals that constitute the mixtures but only the class labels of the sources. We refer to our approach as weak class supervision because, in contrast to other class-based methods such as \cite{Hershey2016}, our method does not require class labels for each time-frequency bin but only a single label for each source constituting the  mixture signal. There are only two requirements: each component in a mixture should have a different source class and the mixtures should have different source class combinations. Intuitively, after observing many mixture signals together with their corresponding source class labels, it becomes possible to predict which class is responsible for which part of the input mixture. Our method makes data labeling easier and enables the use of weakly labeled datasets. It can be useful in many applications with real mixtures where no isolated recordings of the sources are available. For example, it can enable speech separation by only annotating speaker identities, music separation by only annotating active instruments and audio event detection by only weak labels.

Our contributions in this work are as follows:
\begin{itemize}
    \item We propose a model based on deep convolutional $\beta$-VAEs \cite{Higgins2017} that facilitates weak class supervision for source separation.
    \item We show empirically that it is possible to achieve decent separation performance by using only the source class labels for supervision without observing isolated source signals. The separation performance of the proposed method is on par with the performance obtained by signal supervision.
\end{itemize}
 \vspace{-0.3cm}
\section{Background}
The VAE \cite{Kingma2014} is a framework for building probabilistic generative models. Let  $\mathcal{X}=\{\M{x}^{(i)}\}_{i=1}^N$ be the $N$ samples of the training data. We assume that the data are generated by some random process of a latent variable $\M{z}$.
We define a Gaussian prior $p_\theta(\M{z})=\mathcal{N}(\M{z}|\M{0},\M{I})$ on $\M{z}$, an encoder model $q_\phi(\M{z}|\M{x})$ parameterized by $\phi$, and a decoder model $p_\theta(\M{x}|\M{z})$ parameterized by $\theta$. One can use universal function approximators such as neural network models for the encoder-decoder pair. The training dataset $\mathcal{X}$ is used to estimate the model parameters $\phi$ and $\theta$ by maximizing the variational lower bound on the marginal likelihood using stochastic gradient ascent:
\begin{align}
    \mathcal{L}(\theta,\phi \mid \M{x}^{(i)}) &= -D_{KL}(q_\phi(\M{z}|\M{x}^{(i)})||p_\theta(\M{z})) \nonumber \\
    &\qquad + \mathbb{E}_{q_\phi(\M{z}|\M{x}^{(i)})}\left[\log p_\theta(\M{x}^{(i)}|\M{z})\right].
    \label{eq:DKL}
\end{align}
In (\ref{eq:DKL}), $D_{KL}$ denotes Kullback-Leibler (KL) divergence and it can be integrated analytically since we assume both the prior $p_\theta(\M{z})$ and the posterior approximation $q_\phi(\M{z}|\M{x})$ are Gaussian. The KL divergence term acts as a regularizer that enforces the prior $p_\theta(\M{z})$ on the approximate posterior $q_\phi(\M{z}|\M{x})$. The expectation of $\log p_\theta(\M{x}^{(i)}|\M{z})$, which corresponds to a reconstruction error, can be computed with only a single sample from $q_\phi(\M{z}|\M{x}^{(i)})$ as long as the batch size is large enough (e.g. 100). In order to sample from $q_\phi(\M{z}|\M{x}^{(i)})$, we need a reparameterization. We make an isotropic Gaussian assumption such that
\begin{equation}
q_\phi(\M{z}|\M{x}^{(i)})=\mathcal{N}(\M{z}|\M{\mu}^{(i)},\M{\sigma}^{2(i)}\M{I}),
\end{equation}
where $\M{\mu}^{(i)}$ and $\M{\sigma}^{(i)}$ are the outputs of the encoder function. Now, it is possible to use a reparameterization $\M{z}=\M{\mu}+\M{\sigma}\odot\M{\epsilon}$, where $\odot$ denotes element-wise product and $\M{\epsilon} \sim \mathcal{N}(\M{\epsilon}|\M{0},\M{I})$ is an auxiliary noise variable.

Using an isotropic Gaussian for the prior $p_\theta(\M{z})$ encourages independent and disentangled latent units. $\beta$-VAE \cite{Higgins2017} is an extension to the original VAE framework that increases the disentanglement by constraining the latent variable $\M{z}$ further. The only difference of $\beta$-VAE from the original VAE is the $\beta$ coefficient in the lower bound
\begin{align}
    \mathcal{L}_{\beta}(\theta,\phi \mid \M{x}^{(i)},\beta) &= -\beta\, D_{KL}(q_\phi(\M{z}|\M{x}^{(i)})||p_\theta(\M{z})) \nonumber \\
    &\qquad + \mathbb{E}_{q_\phi(\M{z}|\M{x}^{(i)})}\left[\log p_\theta(\M{x}^{(i)}|\M{z})\right],
    \label{eq:bvae}
\end{align}
where $\beta=1$ corresponds to the original formulation and $\beta>1$ restricts the latent variable more while giving less importance to the reconstruction quality.
 \vspace{-0.1cm}
\section{Proposed Model}
\begin{figure*}[!t]
\centering
\subfloat[Source Signal Supervision]{\includegraphics[width=2.3in]{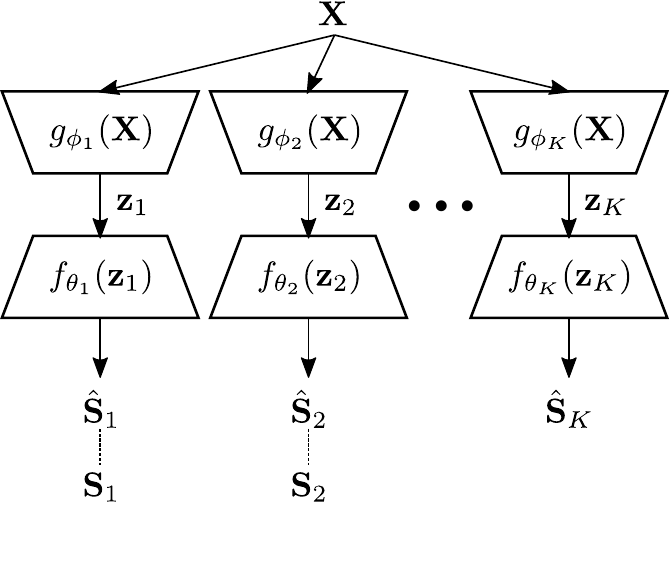}
\label{fig:model_signal_supervision}}
\hfil
\subfloat[Source Class Supervision]{\includegraphics[width=2.3in]{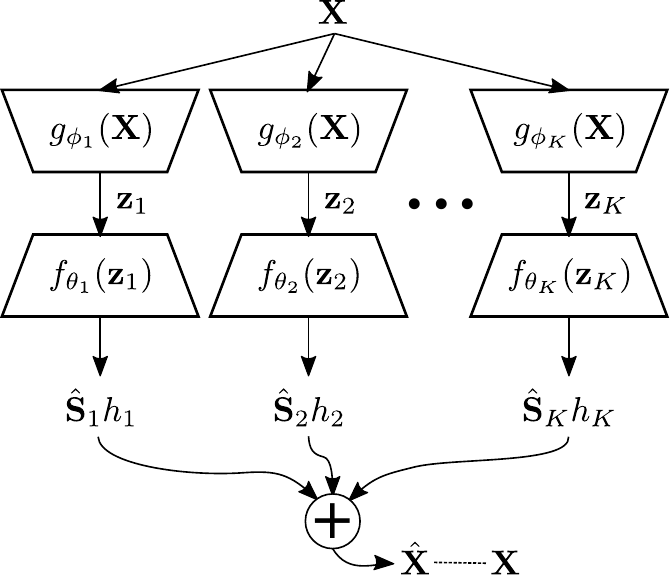}
\label{fig:model_class_supervision}}
\caption{Model architectures for the two different supervision methods. The dotted lines represent the generalized KL divergence loss. The sample index $^{(i)}$ is omitted for simplicity and the mixture $\M{X}$ is assumed to contain sources from classes $1$ and $2$. \protect\subref{fig:model_signal_supervision} In source signal supervision, loss is calculated between the magnitude spectrograms of the estimated sources and the original sources as $D_{GKL}(\M{S}_1||\hat{\M{S}}_1)+D_{GKL}(\M{S}_2||\hat{\M{S}}_2)$. \protect\subref{fig:model_class_supervision} In source class supervision, loss is calculated between the input mixture and the sum of the estimated sources $\hat{\M{S}}_k, k=1,2$ using $D_{GKL}(\M{X}||\hat{\M{X}}=\hat{\M{S}}_1+\hat{\M{S}}_2)$.}
\label{fig_sim}
 \vspace{-0.3cm}
\end{figure*}
 \vspace{-0.1cm}
\subsection{Source Signal Supervision}
\label{sec:signal}
We start by employing autoencoders for the case where we have the source signals for supervision. Magnitude spectrogram of an original source $\M{S}_k^{(i)}$ in a mixture with magnitude spectrogram $\M{X}^{(i)}$ can be estimated by a denoising autoencoder:
\begin{equation}
  \hat{\M{S}}_k^{(i)} = f_{\theta_k}(g_{\phi_k}(\M{X}^{(i)})), \quad f_{\theta_k}(\cdot) \geq 0,
  \label{eq:ae}
\end{equation}
where $g_{\phi_k}(\cdot)$ and $f_{\theta_k}(\cdot)$ are the neural network encoder-decoder pair for the $k$-th source class, respectively. Let $\mathcal{X}_s=\{(\M{X}^{(i)}, \M{S}^{(i)})\}_{i=1}^N$ be the training dataset where $\M{S}^{(i)}$ contains the magnitude spectrograms of the original sources corresponding to the mixture $\M{X}^{(i)}$. Then, we can arrive at the approach in \cite{Grais2017} with the exception of using the generalized KL divergence \cite{Lee2001} instead of the squared error:
\begin{equation}
    D_{GKL}(\M{S}_k^{(i)}||\hat{\M{S}}_k^{(i)}) = \sum_{t,f} \M{S}_{k_{t,f}}^{(i)} \log{\frac{\M{S}_{k_{t,f}}^{(i)}}{\hat{\M{S}}_{k_{t,f}}^{(i)}}} - \M{S}_{k_{t,f}}^{(i)} + \hat{\M{S}}_{k_{t,f}}^{(i)},
    \label{eq:DGKL}
\end{equation}
where ${t,f}$ are the indices for the time and frequency bins of the $k$th source $\M{S}_k^{(i)}$, respectively. The model architecture for source signal supervision is given in Fig. \ref{fig:model_signal_supervision}.
 \vspace{-0.2cm}
\subsection{Source Class Supervision}
\label{sec:class}
We now turn our attention to employing autoencoders for the case where we do not have access to the source signals for supervision but only their classes. Assume that we have a training dataset such that $\mathcal{X}_h=\{(\M{X}^{(i)}, \M{h}^{(i)})\}_{i=1}^N$ where $h^{(i)}_k=1$ if the mixture contains source $k$, and $h^{(i)}_k=0$, otherwise. Here we introduce our non-negative model which will allow estimating the sources $\hat{\M{S}}_k^{(i)}$ in (\ref{eq:ae}) by using only the source classes $h_k^{(i)}$ instead of the original sources $\M{S}_k^{(i)}$:
\begin{equation}
  \hat{\M{X}}^{(i)} = \sum_{k=1}^K \hat{\M{S}}_k^{(i)} h_k^{(i)}.
  \label{eq:ae_sum}
\end{equation}
In order to find the source estimates $\hat{\M{S}}_k^{(i)}$, we use the loss function $D_{GKL}(\M{X}^{(i)}||\hat{\M{X}}^{(i)})$, therefore we aim to reconstruct the mixture $\M{X}^{(i)}$ while using only the autoencoders associated with their source classes. While training, the model needs to observe mixtures when some classes exist, and some classes are absent. This requires having different classes in the mixtures. The model architecture for source class supervision is given in Fig. \ref{fig:model_class_supervision}.

The resulting non-negative model is similar to NMF in the sense that, we use additive components to decompose the mixture. The difference is that, in NMF, one uses linear templates for representing the components, whereas our model is able to use non-linear and more expressive neural network models. In our model, each encoder learns to ignore the noise related to the other sources in the mixture and only learn an embedding of the source class associated with that autoencoder, while the decoder learns to reconstruct the associated source signal.
 \vspace{-0.2cm}
\subsection{Incorporating VAEs}
\label{sec:vae}
Until now, we developed our model based on standard autoencoders. We now propose utilizing the $\beta$-VAE framework that is given in (\ref{eq:bvae}). 
We assume a Poisson distribution on the magnitude mixture spectrogram  $\M{X}^{(i)}$  with a mean parameter $\hat{\M{X}}^{(i)}$ given in (\ref{eq:ae_sum}):
\begin{equation}
    p_\theta(\M{X}^{(i)}|\M{z}) = \mathcal{PO}(\M{X}^{(i)}|\M{\lambda}=\hat{\M{X}}^{(i)}).
    \label{eq:poisson}
\end{equation}
A Poisson distribution is commonly employed to model magnitude spectrograms (e.g. in NMF models) and corresponds to using the unnormalized KL divergence \cite{Cemgil2009}. To train the generative model in (\ref{eq:poisson}), we convert (\ref{eq:bvae}) to a loss function
\begin{align}
    L(\cdot) &= -\beta \frac{1}{2} \sum_{k,j}\left(1 + \log((\sigma_{k,j}^{(i)})^2) - (\mu_{k,j}^{(i)})^2 - (\sigma_{k,j}^{(i)})^2 \right) \nonumber \\
    &\qquad + D_{GKL}(\M{X}^{(i)}||\hat{\M{X}}^{(i)}),
    \label{eq:loss}
\end{align}
where $k$ and $j$ denotes class and latent unit indices, respectively. The first term, a KL divergence, is derived based on the assumption that both $p_\theta(\M{z})$ and $q_\phi(\M{z}|\M{x})$ are Gaussian. This term can be viewed as a prior constraint on the latent units, which encourages independence not only between the latent units but also between the sources. Therefore, it provides a useful regularization effect for this problem. Due to the fact that VAEs are generative models and we assign a VAE to each class, we are able to obtain a generative signal model $f_{\theta_k}(\M{z}_k)$ for each class $k$ without observing the source signals. We can draw a sample from the decoder $f_{\theta_k}(\M{z}_k)$ by supplying a $\M{z}_k \sim \mathcal{N}(\M{z}_k|\M{0},\M{I})$.
 \vspace{-0.2cm}
\subsection{Encoder-Decoder Network Architecture}
\label{sec:arch}
We use Convolutional Neural Networks (CNN) similar to the one proposed in \cite{Hsu2017}. It is designed to work with magnitude spectrograms of size $T \times F$ where $T$ and $F$ represent the size of the time and frequency axes, respectively. Our encoder architecture is given in Table \ref{tab:architecture} where Conv refers to convolutional layers, FC refers to the fully-connected layer and Gauss refers to the Gaussian latent output layer. The first layer learns templates along the frequency axis similar to NMF while the next two layers focus on temporal patterns. The encoder-decoder networks have a symmetrical structure and the decoder uses transposed convolutions. We use rectified linear unit (ReLU) non-linearities for all layers except for the last layers of the encoder and the decoder. Batch normalization is similarly applied to every layer except for the last layers. We use the softplus function $o(x)=\log (1+e^x)$ in the last layer of the decoder to produce a non-negative output.

\begin{table}[t]
  \caption{Encoder Network Architecture}
  \label{tab:architecture}
  \centering
  \begin{tabular}{ | l | c | c | c | c | c | }
    \hline
     & Conv1 & Conv2 & Conv3 & FC & Gauss \\ \hline
    Filters/units & 128 & 128 & 256 & 512 & 128 \\ \hline
    Filter size & 1$\times F$ & 4$\times$1 & 4$\times$1 & - & - \\ \hline
    Stride & 1$\times$1 & 2$\times$1 & 2$\times$1 & - & - \\
    \hline
  \end{tabular}
   \vspace{-0.4cm}
\end{table}

\section{Experiments}
\begin{figure*}
  \centering
  \includegraphics[width=0.85\linewidth]{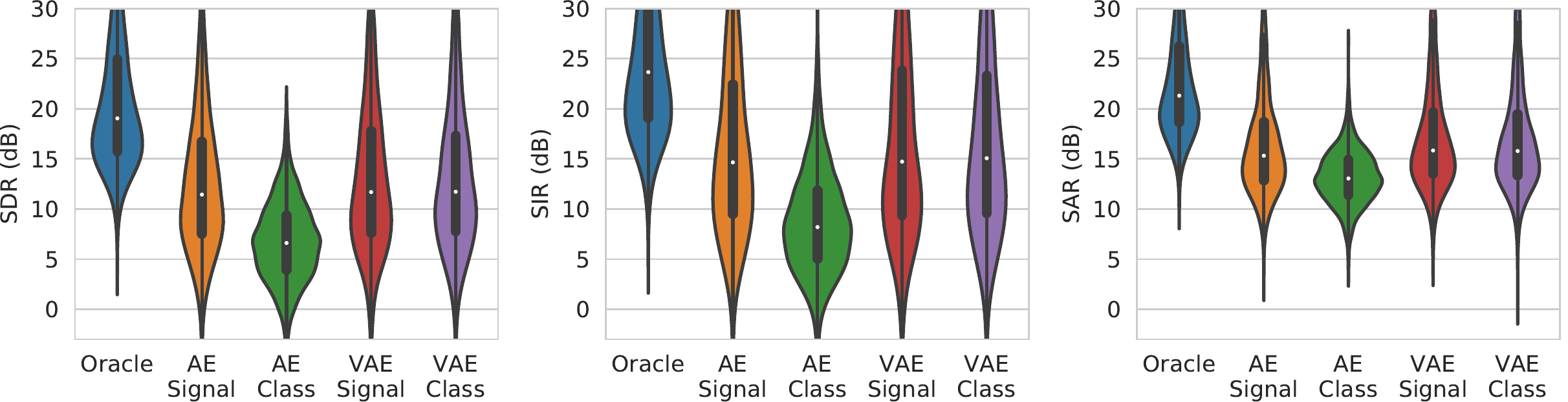}
    \caption{The distributions of Source to Distortion Ratio (SDR), Source to Interference Ratio (SIR), and Source to Artifacts Ratio (SAR) metrics are shown as violin plots. AE and VAE refer to the autoencoder model and the variational autoencoder model, respectively. Signal and Class refer to source signal supervision and source class supervision, respectively.}
     \vspace{-0.5cm}
    \label{fig:compare_ae_vae_source_label}
\end{figure*}

\begin{figure}
  \centering
  \includegraphics[width=0.72\linewidth]{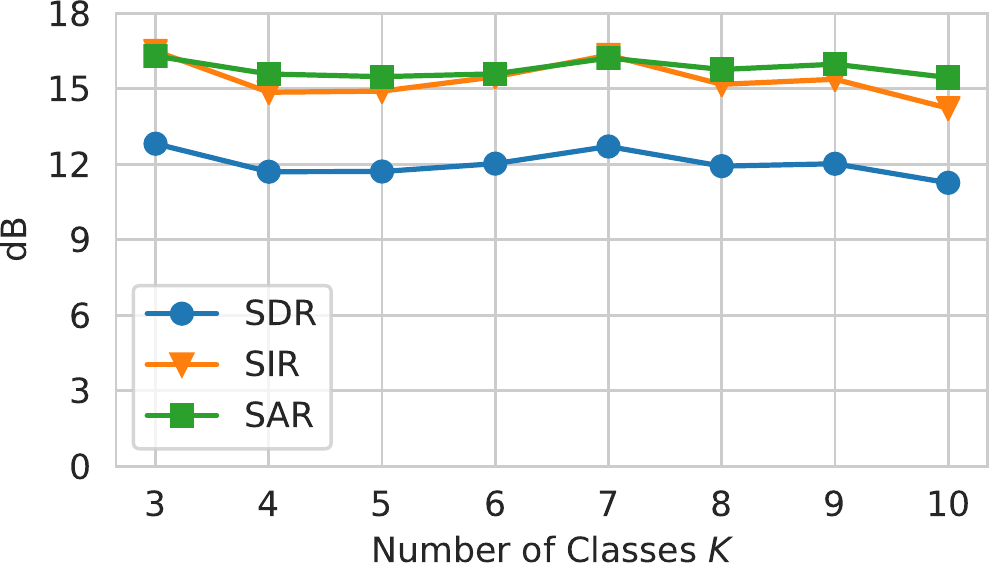}
  \vspace{-0.2cm}
    \caption{Source separation performance with respect to number of classes. The medians of the metrics are given.}
    \label{fig:vary_num_classes}
    \vspace{-0.3cm}
\end{figure}

\subsection{Dataset}
We evaluate our models on the mixtures we generate from the Speech Commands Dataset (SCD) \cite{Warden2018}. The dataset contains one-second recordings of utterances. This is a challenging dataset because there are many different speakers, recording conditions, loudness, alignments, noise, mislabeling etc. We select a subset that contains digit utterances and use it as a 10-class source dataset. We preprocess the recordings by first downsampling to $8$ kHz and then normalizing the Root Mean Square (RMS) value of each recording to the mean RMS value of the training set. This RMS normalization is needed because the loudness of the recordings vary widely.

We define the set of classes which a mixture dataset contain as $\mathcal{C}_{a:b}=\{i\in\mathbb{Z}\mid a\leq i< b\}$ where $a\geq 0$ and $3\leq b\leq 10$. We include all combinations of the classes $\mathcal{C}_{a:b}$ in the mixture datasets we generate, therefore the number of combinations are $C(K,O)=\frac{K!}{O!(K-O)!}$ where $K$ is the number of classes in dataset and $O=\sum_k h_k$ is the number of components in mixtures.
For example, using the classes $\mathcal{C}_{6:9}$ with $K=3$ classes and $O=2$ components will produce $C(3,2)=3$ class label combinations: $\{6,7\}$, $\{6,8\}$ and $\{7,8\}$. We constrain our experiments to two-component ($O=2$) mixtures but this method is also applicable for more components. For the two-component case, we need a minimum of 3 classes because our model requires observing mixtures with different class combinations. Otherwise, there will be only a single combination.

To create a mixture dataset, we first generate a list of all class label combinations from $\mathcal{C}_{a:b}$ and cycle through the list while choosing recordings randomly for each of the labels in the combination, then mix the chosen recordings at $-6, 0$ and $6$ dB. 
We generate $15000$ training, $1875$ validation, and $1875$ testing samples where each mixture partition is generated from the respective partition of the SCD. The generated mixture dataset contains almost every recording in the SCD once. Also, the number of mixtures for each label combination is roughly the same. Finally, we calculate a $512$-point (64 ms) Short Time Fourier Transform (STFT) of the mixtures using a Hann window with $50\%$ overlap.
\subsection{Experimental Setup}
We reconstruct the separated sources in the time domain using a soft time-frequency mask (Wiener filter) and the mixture phase:
\begin{equation}
    \hat{\M{y}}_t^{(i)} = \mathrm{STFT}^{-1} \left(\frac{(\hat{\M{S}}_t^{(i)})^2}{\sum_c (\hat{\M{S}}_c^{(i)})^2} \odot \M{X}^{(i)} \odot e^{i\M{\Phi}} \right)
\end{equation}
where $t$ is the target component index, $c$ is a component index, $\M{\Phi}$ is the phase of the mixture spectrogram and $\mathrm{STFT}^{-1}$ is the inverse STFT. All the operations are element-wise.

The evaluation is based on the BSS\_EVAL metrics \cite{Vincent2006}. We use PyTorch 1.0.1 with an NVIDIA GeForce GTX 1080 Ti. The source code and a demo page are available online\footnote{\url{https://github.com/ertug/Weak_Class_Source_Separation}}. We use the Adam optimization algorithm with default parameters \cite{Kingma2015}. We set $\beta=10$ and the batch size is $100$. We evaluate the reconstruction loss on the validation set at every 200 iterations and stop training if it does not improve for 10 evaluations. We use the model with the best reconstruction loss. We obtain reasonable results without much hyperparameter tuning.

We assume that the test set $\mathcal{X}_{test}=\{(\M{x}^{(i)}, \M{h}^{(i)})\}_{i=1}^N$ contains the source class labels $\M{h}^{(i)}$ as the training set so that we can activate the associated pre-trained autoencoders to solve the two-component source separation problems in the testing stage. This assumption is not much of a weakness of the approach because one can easily annotate a mixture with only class labels and let the system do the harder task of source separation. Also, it can be readily addressed by using a classifier that predicts the class labels present in the mixture.
\subsection{Results}
Our first experiment compares the AE model \cite{Grais2017} to the VAE model introduced in Section \ref{sec:vae} and  signal supervision to class supervision. AE model with signal supervision and class supervision are described in Sections \ref{sec:signal} and \ref{sec:class}, respectively. Both the AE and VAE models use the same network architecture as given in Section \ref{sec:arch}. We train and evaluate the models on a mixture dataset containing all of the 10 classes, i.e. $\mathcal{C}_{0:10}$. We also include oracle metrics where the magnitude spectrograms of the original sources are used. We report the results for the mixtures obtained at 0 dB, since the performance gap among different interference levels (-6, 0, 6 dBs) was similar for all methods. The results are given in terms of violin plot in Fig. \ref{fig:compare_ae_vae_source_label}. The white dot represents the median value and the thick line represents the inter-quartile range for each method. Our baseline AE model is able to demonstrate a decent separation performance when supervised by source signals but fails when supervised by only source classes. In contrast, the VAE model is able to perform well, even when supervised by only source classes. This improvement in performance shows that the proposed VAE model can be a powerful algorithm for real-world mixtures. At the same time, the variance in SDR, SIR and SAR values obtained by the proposed method is similar to the variance values obtained by the signal supervised AE as it is seen from the inter-quartile ranges. 

In our second experiment, we vary the number of source classes in the dataset for the class supervised VAE model in order to see if it has an effect on the performance. For each number of classes $3\leq K\leq 10$, we train and evaluate the model on a mixture dataset containing the classes $\mathcal{C}_{0:K}$. Only for the $K=3$ case, we train and evaluate the model on 3 different mixture datasets containing disjoint classes $\mathcal{C}_{0:3}$, $\mathcal{C}_{3:6}$ and $\mathcal{C}_{6:9}$. This is to reduce the variance of metrics due to inter-class differences but we also observed that the variance is already small. The results are given in Fig. \ref{fig:vary_num_classes}. The number of classes does not seem to have a significant effect.

\section{Conclusion}
We propose a non-negative VAE model for source separation that is useful when we do not have access to the isolated source signals but only the class labels. Our experiments demonstrate that class supervision performs as good as signal supervision. Our future work will include using a multi-label classifier to predict $\M{h}^{(i)}$, so that observing the class labels in the testing stage will not be required. Finally, this work has the potential to be extended to other areas where NMF is already employed.

\bibliographystyle{IEEEtran}
\bibliography{IEEEabrv,refs}

\end{document}